\documentclass[10pt,prx,twocolumn,amsmath,amssymb,floatfix,notitlepage]{revtex4-2}
\usepackage[T1]{fontenc}
\usepackage[utf8]{inputenc}
\usepackage{lmodern}
\usepackage{microtype,bm,bbm}
\usepackage{graphicx,booktabs}
\usepackage{times}
\usepackage[usenames,dvipsnames]{xcolor}
\usepackage[english]{babel}
\usepackage{upgreek}
\usepackage{braket}
\usepackage{subfigure}
\usepackage{dsfont}
\usepackage{transparent}
\usepackage{layouts}

\usepackage{hyperref}
\hypersetup{
  colorlinks,
  allcolors=NavyBlue,
  linktoc=all
}

\usepackage[capitalize]{cleveref}
\crefformat{section}{Sec.~#2#1#3}

\newcommand{\id}{\mathbbm{1}}

\renewcommand*\vec[1]{\mathbf{#1}}

\graphicspath{ {figures/}}
\graphicspath{ {inkscape_figures/}}

\begin{document}

\title{Quantum Simulation of Spin-1 XXZ-Heisenberg Models and the Haldane Phase with Dysprosium}
\author{Katharina Brechtelsbauer}
\author{Johannes Mögerle}
\author{Hans Peter Büchler}
\affiliation{Institute for Theoretical Physics III and Center for Integrated Quantum Science and Technology, University of Stuttgart, Pfaffenwaldring 57, 70569 Stuttgart, Germany }
\date{\today}

\begin{abstract}
Dysprosium atoms have proven to be a promising platform for quantum simulation due to their strong magnetic moment and high tunability of interactions.
In this work, we propose Dysprosium atoms for simulating the one-dimensional spin-1 XXZ-Heisenberg model, which is known to have a rich phase diagram including the famous Haldane phase. 
To realize the model, we make use of the strong dipolar exchange interactions that naturally occur in the ground state of Dysprosium due to its large electron angular momentum of J=8.
To implement spin-1 particles, we encode the spin degree of freedom into three Zeeman sublevels, which are energetically isolated by applying a magnetic field. Using the density-matrix renormalization group, we analyze the ground-state properties of the resulting effective model. We find that a chain of fermionic Dysprosium atoms in a suitable magnetic field can form a Haldane state with the characteristic spin-1/2 edge modes. Furthermore, we discuss the use of AC Stark shifts and Raman-type schemes for bosonic Dysprosium to isolate effective spin-1 systems and to increase the tunability of model parameters.

\end{abstract}
\maketitle

\section{Introduction}

Topological phases of matter can exhibit exciting properties such as long-range entanglement, gapless edge states, robust ground state degeneracy, or excitations that obey anionic statistics or carry fractionalized charges \cite{Wen2017}. A multitude of theoretical proposals \cite{Baranov2005,Sorensen2005,Micheli2006,Cooper2013,Yao2013,Gerster2017,Andrews2018,Verresen2021,Weber2022} together with the great experimental progress in the recent years enables to probe these phenomena on platforms like cold atoms, Rydberg systems, polar molecules or trapped ions  \cite{Altman2021,Gross2017,Browaeys2020,Monroe2021,Semeghini2021}. While topological order cannot appear in one-dimensional bosonic systems, topological and trivial phases can still be distinguished applying the concept of symmetry protected topological phases \cite{Chen2011,Chen2013}. A famous example for a symmetry protected topological phase is the Haldane phase \cite{Haldane1983}, which is the predicted ground state for the anti-ferromagnetic spin-1 Heisenberg model. One fascinating property of the Haldane phase is that the integer spins at the edge fractionalize into spin-1/2 degrees of freedom. 
While the Haldane phase has already been observed in systems composed by spin-1/2 particles \cite{Leseleuc2019,Sompet2022}, here we propose Dysprosium atoms to directly implement spin-1 particles and observe the Haldane phase and the fractionalization of the edges.

Highly magnetic atoms like Dysprosium, Erbium or Chromium stand out due to their strong magnetic moments \cite{Chomaz2023}. Compared with the usual alkali atoms the magnetic dipole-dipole interactions are up to 100 times stronger. Furthermore, among the naturally abundant isotopes of Dysprosium, Erbium, and Chromium there are fermionic atoms as well as bosonic, which come along with a variety of different features like variable scattering properties. All this makes them an extremely versatile platform for quantum simulation, which is currently utilized in many experiments. For example, Bose-Einstein condensates of Dysprosium and Erbium are used for investigating quantum droplets \cite{Kadau2016,Chomaz2016}  and observing supersolid states of matter \cite{Boettcher2019,Chomaz2019,Tanzi2019}; for more details see \cite{Boettcher2021,Chomaz2023}. Furthermore, great progress has been made in the preparation of specific Zeemann sublevels \cite{Claude2024,Douglas2024}, allowing the observation of spin squeezing \cite{Douglas2024}. Recent experiments combine magnetic atoms and optical lattices aiming for the simulation of extended Hubbard models and quantum magnetism \cite{Paz2013,Paz2016,Baier2016,Fersterer2019,Lepoutre2019,Patscheider2020,Hughes2023,Bilinskaya2024}. Typically, the lattice constants lie between 500 nm and 200 nm, resulting in strong magnetic dipole-dipole interactions. For deep lattices, where tunneling between lattice sites is strongly suppressed, Mott-insulating states with one particle per lattice site can be prepared. In contrast to typical cold atom experiments \cite{Gross2017}, superexchange interactions are negligible and the resulting spin model in this limit is dominated by long-range dipole-dipole interactions. In the presence of strong magnetic fields, where magnetisation-changing collisions can be ignored, the dynamics of these systems is then described by a XXZ-Heisenberg model \cite{Lepoutre2019,Patscheider2020}.

\begin{figure}[tb]
\includegraphics[width= \linewidth]{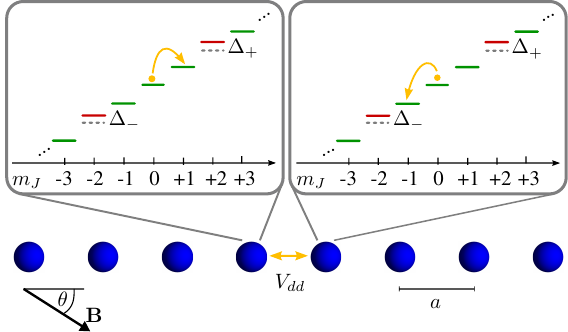}
\caption{Setup for the realization of the Haldane phase with Dysprosium. Dysprosium atoms are trapped in an optical lattice within a magnetic field $\mathbf{B}$. The ground state of Dysprosium consists of 17 magnetic sublevels. Exchange interactions allow for processes where the $m_J$ quantum number of one atom is increased by one while the $m_J$ quantum number of another atom is lowered by one. In particular, it can happen that  at least one atom is transferred to a state with quantum number $m_J\neq0,\pm1$. However, if the states with quantum number $m_J=\pm2$ are detuned, these transitions are energetically forbidden and the system behaves effectively like a chain of spin-1 particles.}
\label{fig:detuningsandswt}
\end{figure}

In this work we propose Dysprosium atoms for the realization of the spin-1 XXZ-Heisenberg model and especially the Haldane phase. We assume the atoms to be trapped in a deep optical lattice within a strong magnetic field such that the effective Hamiltonian of the system is given by the XXZ-Heisenberg Hamiltonian. In its ground state bosonic Dysprosium has a total angular momentum of 8. Thus, to mimic spin-1 particles the levels with magnetic quantum numbers $m_J=\pm2$ have to be shifted out of resonance, such that transitions out of the Spin-1 subspace are energetically forbidden (see \cref{fig:detuningsandswt}). We show that this can be achieved by either applying stark shifts or utilizing the hyperfine coupling in fermionic Dysprosium. Using density matrix renormalization group techniques (DMRG) we analyze the resulting effective model and show that for a suitable set of parameters the ground state of the system lies indeed in the Haldane phase, where fractionalized edge modes can be observed. Furthermore, we use the Schrieffer-Wolff transformation to take into account corrections that arise from higher order processes where the energetically forbidden states are virtually occupied. Our system differs from other proposals \cite{Cohen2014,Cohen2015,Senko2015,Moegerle2024} by an anti-ferromagnetic static coupling term. While for short-range interactions the Haldane phase does not survive in the presence of such a term \cite{Schulz1986}, it is a special property of systems with dipole-dipole interactions that the Haldane phase is existent even for small anti-ferromagnetic static interactions \cite{Gong2016}. Our work opens the possibility to study this shift of the phase boundaries experimentally.

The rest of this work is structured as follows. First, we present the general setup and the effective model that can be simulated with Dysprosium in \cref{sec:system}. In \cref{sec:phasediagram} we then review the ground state phase diagram of this model. Finally, a detailed discussion of possible realizations of Spin-1 particles with Dysprosium is given in \cref{sec:isolation-of-spin-1-subsystem}. Furthermore, we analyze the effect of perturbations that come along with the implementation.

\section{System}
\label{sec:system}

We consider a one dimensional chain of ground state Dysprosium atoms, which are trapped in an optical lattice with lattice constant $a$ and one atom per lattice site. 
The atoms interact via magnetic dipole-dipole interactions 
\begin{align}
V_{dd}&=\frac{\mu_0}{4 \pi}  \sum_{ij} \frac{\boldsymbol{\mu}_i \cdot \boldsymbol{\mu}_j}{r_{ij}^3} -3  \frac{(\boldsymbol{\mu}_i \cdot \mathbf{r}_{ij})(\boldsymbol{\mu}_j \cdot \mathbf{r}_{ij})}{r_{ij}^5},
\label{eq:dipoledipole}
\end{align}

where, $r_{ij}=a|i-j|$ is the distance between the $i$'th and $j$'th atom and $\boldsymbol{\mu}_i=\mu_B g_J \mathbf{J}_i$ the magnetic moment, with $\mu_B$ the Bohr magneton $g_J$ the Lande factor and $\mathbf{J}_i$ the angular momentum operator acting on the $i$'th electron.  A strong magnetic field $\mathbf{B}$ ensures that the total angular momentum $\mathbf{J}_\text{tot}$ is conserved. Without loss of generality we choose the magnetic field to point into the $z$-direction.  

In its ground state dysprosium has an angular momentum quantum number of $J=8$ and the different ground states can be labelled by the eigenvalues $m_J$ of the z-component of the angular momentum.

We assume that the states $\ket{m_J=0}$  and $\ket{\pm1}$ can be energetically isolated such that we effectively obtain a system composed by spin-1 particles. 
Within a rotating frame the dynamics of the system is then described by the anisotropic XXZ-Heisenberg Hamiltonian 
\begin{align}
 H_\text{eff} &= \sum_{i < j} \frac{1}{\left|i-j\right|^3}
                \left( \frac{J_{xy}}{2} ( S^+_i S^-_j + S^-_i S^+_j ) + J_{z} S^z_i S^z_j  \right) \notag \\
            &+ \sum_i D (S^z_i)^2.
            \label{eq:heisenberg}
\end{align} 

Here, $\mathbf{S}_i$ are spin-1 operators and the couplings are $J_{xy}=-\mu_0 d^2/(8 \pi  a^3) (1-3 \cos^2(\theta))$ and $J_z=\mu_0 m^2 /(4 \pi  a^3)(1-3 \cos^2(\theta))$, where $\theta$ is the angle between the chain and the magnetic field, $d=\sqrt{2}\bra{+1} \mu^x \ket{0}$ is the transition dipole moment and $m=\bra{+1} \mu^z \ket{+1}$ the permanent magnetic moment. Note that for Dysprosium with angular momentum $J=8$ the exchange coupling $J_{xy}$ is about 36 times larger than the exchange coupling one would expect for a natural spin one system. The single-site anisotropy $D$ originates from the assumption that the transitions from $\ket{0}$ to $\ket{\pm 1}$ are slightly detuned by $D=( \epsilon_{-1}+\epsilon_{+1}-2 \epsilon_0)/2$. Here, $\epsilon_{m_J}$ is the energy of the state $\ket{m_J}$. For bosonic dysprosium atoms placed in a magnetic field with strength $B$ the energies of the magnetic states are simply $\epsilon_{m_J}=\mu_B B g_J m_J$ and the (uniaxal single-ion) anisotropy is $D=0$. 

Before we discuss in \cref{sec:isolation-of-spin-1-subsystem} how the states $\ket{0}$  and $\ket{\pm1}$ can be isolated, we briefly consider the ground state phase diagram of the effective Hamiltonian in \cref{eq:heisenberg} in \cref{sec:phasediagram}.

\section{Phase Diagram}
\label{sec:phasediagram}

\begin{figure}[t]
\includegraphics[width= \linewidth]{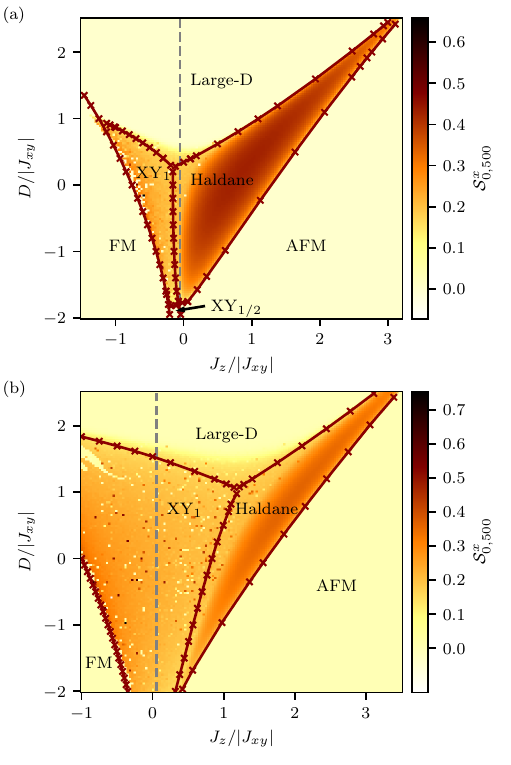}
\caption{Phase diagram of the Hamiltonian in \cref{eq:heisenberg} for $J_{xy} > 0$ (a) and $J_{xy} < 0$ (b) . The color map illustrates the string order parameter $\mathcal{S}^x_{0,500}$. It has been determined using infinite DMRG calculations with bond dimension 400. In the limit $|i-j| \rightarrow \infty$ the string order $\mathcal{S}^x_{ij}$ is expected to be zero everywhere except in the Haldane phase. However, in the XY phase it decays slowly causing non-zero values even for a distance of $j-i=500$ \cite{Su2012}. The straight lines represent the phase boundaries. For the XY-Haldane transition the boundaries suggested from the string order deviate strongly from the actual boundaries, since in the XY-phase the results of the DMRG-calculations depend strongly on the bond dimension. The dashed line indicates the regime that can be simulated using bare fermionic dysprosium. Applying Raman-type schemes enables to tune $J_z$ and thus to cover a wider range of the phase diagram.}
\label{fig:phasediagram}
\end{figure}

In \cref{fig:phasediagram} we show the phase diagram of the Hamiltonian in \cref{eq:heisenberg}. The ground state properties are evaluated with infinite DMRG calculations using TeNPy \cite{Hauschild2018}. To determine the phase boundaries we furthermore use repeated application of finite DMRG, which allows us to also compute excitations. A detailed discussion of the analysis of the critical lines is given in \cref{ap:boundaries}.

Similar to the phase diagram of the anisotropic spin-1 XXZ-Heisenberg Hamiltonian with nearest neighbor interactions  \cite{Schulz1986,Kennedy1992,Chen2003}  our phase diagrams include the Haldane phase, the large-D phase, the Ising ferromagnetic (FM) and the Ising anti-ferromagnetic phase (AFM) as well as two (anti-)ferromagnetic XY phases with quasi long-range order (XY$_{1}$ and XY$_{1/2}$). However, due to the long-range character of the interactions in \cref{eq:heisenberg}  the phase boundaries change \cite{Gong2016}. In particular, the critical line separating the Haldane- and XY-phase is shifted to negative $J_z$ for the anti-ferromagnetic model ($J_{xy}>0$) and to positive  $J_z$ for the ferromagnetic model ($J_{xy}<0$). In the following we discuss the characteristics of the different phases mentioned before. 

We start with the limits of large $|J_z|$. For large positive $J_z$ the energy is minimized if neighboring spins point in opposite directions and the system is in the Ising anti-ferromagnetic phase. The states where all spins are in the state $\ket{+1}$ or $\ket{-1}$ are always eigenstates of the Hamiltonian in \cref{eq:heisenberg} with energy $E_\text{FM}=\sum_{i>j}J_z/(i-j)^3+ND$. For large negative $J_z$ this energy becomes the smallest energy and the system is in the Ising ferromagnetic phase. Both phases, the Ising anti-ferromagnetic and the Ising ferromagnetic phase, are gapped and have long range order indicated by non-vanishing correlations $ \langle(-1)^{|i-j|} S_i^z S_j^z \rangle$ and $\langle S_i^z S_j^z \rangle$ respectively. 
 
For large positive $D$ the term proportional to $S_z^2$ dominates, and the smallest energy state is the one where all spins are in the state $\ket{0}$. The so called large-D phase is thus topologically trivial and disordered and has a gapped ground state. Vice versa, in the limit of large negative $D$  any state where one atom is in $\ket{0}$ has a much higher energy than states that only contain $\ket{\pm1}$ and the system effectively maps to a spin $1/2$-XXZ Heisenberg model with exchange couplings $\propto J_{xy}^2/D$ \cite{Chen2003}. In this limit one thus finds, depending on the sign of $J_{xy}$, the gapless anti-ferromagnetic or ferromagnetic spin-1/2 XY phase \cite{Giamarchi2007,Maghrebi2017}. 
The ground state in this phase is characterized by quasi long-range order, indicated by powerlaw correlations $ \langle (S_i^+)^2 (S_j^-)^2 \rangle$ and $ \langle (-1)^{|i-j|} (S_i^+)^2 (S_j^-)^2 \rangle$, respectively, as quantum fluctuations prevent long-range order in one dimension. Due to the form of the exchange couplings, we expect the width of then spin-1/2 XY phase to decrease with $1/|D|$ \cite{Schulz1986,Chen2003}.

Finally, the competition of $D$,$J_{xy}$ and $J_z$ gives rise to 2 more phases. The first one is the (anti-)ferromagnetic spin-1 XY phase with quasi long-range order, powerlaw correlations $\langle S_i^+ S_j^- \rangle$ or $\langle (-1)^{|i-j|} S_i^+ S_j^- \rangle$ and gapless excitations. 
The second phase is the famous Haldane phase. The Haldane phase is a symmetry protected topological phase with hidden order. It can be characterized by non-vanishing string order parameters $\mathcal{S}_{ij}^\alpha=\langle S_i^\alpha \prod_{i<k<j} (-1)^{S_k^\alpha} S_j^\alpha \rangle $ with $\alpha=x,y,z$. The string order $\mathcal{S}^z_{ij}$ is non-zero in the Ising ferromagnetic and anti-ferromagnetic phase as well, but vanishes in the $XY$ phases and the Large-D phase. The string order $\mathcal{S}^x_{ij}$ vanishes in the $XY_{1/2}$ phase, the Large-D phase, the Ising ferromagnetic, and the Ising anti-ferromagnetic phase, but is finite in the $XY_1$ phase.  This is caused by the fact that the string order decays slowly in the $XY_1$ phase, such that even for large finite distances non-zero values are obtained \cite{Su2012}. 

While the string order $\mathcal{S}^z_{ij}$ is non-zero in the Ising anti-ferromagnetic phase as well, the string order $\mathcal{S}^x_{ij}$ is nonzero only in the Haldane phase. In figure \cref{fig:phasediagram} we show the strength of the string order $\mathcal{S}^x_{ij}$ for different values of $J_z$ and $D$. Other than expected we also obtain non-zero values in the XY$_1$ phase and near the phase boundaries of the large-D phase. This is caused by the fact that the string order decays slowly in this regions, such that even for large finite distances non-zero values are obtained \cite{Su2012}. 

Another feature of the Haldane phase is the occurrence of spin-1/2 edge excitations. Deep in the Haldane phase the spin-1/2 edge states give rise to a fourfold degeneracy of the ground state. In finite systems, the degeneracy is slightly lifted due to the non-zero overlap of the edge states.

There are several symmetries that protect the Haldane phase \cite{Pollmann2010}: time-reversal symmetry and the dihedral symmetry D$_2$ (rotations by $\pi$). Here, we have both. Note, if only time-reversal symmetry is present the string order can vanish.

\section{Isolation of spin 1 subsystem}
\label{sec:isolation-of-spin-1-subsystem}

The dipole-dipole coupling in \cref{eq:dipoledipole} give rise to exchange interactions where the $m_J$ quantum number of one atom is increased by one while the $m_J$ quantum number of another atom is lowered by one (see \cref{fig:detuningsandswt}). For bosonic Dysprosium atoms in a magnetic field all energy levels of the ground state manifold are equidistant such that all processes caused by exchange interactions are resonant. As a consequence, after some time states with quantum numbers $|m_J|\geq 2$ will be occupied - in other words the atoms obviously do not behave like spin-1 particles. To prevent this the states with quantum numbers $m_J=\pm2$ have to be detuned such that the transitions to these states are energetically forbidden.  

To illustrate this we consider two Dysprosium atoms both prepared in any of the states $\ket{0}$ and $\ket{\pm1}$. If both atoms are prepared in the state $\ket{+1}$ the exchange interactions drive the transition to the state $\ket{+2} \otimes \ket{0}$. The energy of the final state differs from the energy of the initial state by $\Delta_+$ , where $\Delta_+$ is the detuning of the state $\ket{+ 2}$ (see \cref{fig:detuningsandswt}). The energy difference of the states $\ket{+ 2} \otimes \ket{-1}$ and $\ket{+ 1} \otimes \ket{0}$ is also given by the detunings $\Delta_+$. Analogously, unwanted transions that involve $\ket{-2}$ have an energy difference of $\Delta_-$,  with $\Delta_-$ the detuning of the state $\ket{-2}$. Furthermore, for the initial state $\ket{+1} \otimes \ket{- 1}$ the unwanted transition to $\ket{+2} \otimes \ket{-2}$ is detuned by $\Delta_++\Delta_-$. To avoid any transition out of the spin-1 subspace the energy costs of all the above discussed transitions have to be much larger than the strength of the dipole-dipole interaction. In conclusion, Dysprosium atoms can mimic spin-1 particles if the states $\ket{\pm 2}$ are detuned by $\Delta_\pm$, where the detunings satisfy 
$|\Delta_+|,|\Delta_-|, |\Delta_+ +\Delta_-|\gg \text{max}(|V_{dd}|) \propto J_{xy}$.

In the following we discuss two possibilities for detuning the states $\ket{\pm 2}$. First, we demonstrate that for fermionic Dysprosium the requested conditions on the detunings $\Delta_\pm$ can be fulfilled due to hyperfine coupling. Afterwards, we investigate the usage of AC-stark shifts and Raman-type schemes for bosonic atoms. 

\subsection{Fermionic Dysprosium}

\subsubsection{Realization of Spin-1 systems}

Fermionic Dysprosium has a nuclear spin of $I=5/2$. The coupling between the nuclear spin and the total angular momentum of the electrons is described by

\begin{align}
H_\text{HFS}=h \left(A\  \mathbf{I}\mathbf{J} +Q\ \frac{3(\mathbf{I}\mathbf{J})^2+\frac{3}{2}\mathbf{I}\mathbf{J}-I(I+1)J(J+1)}{2I(2I-1)J(2J-1)}\right),
\label{eq:Hhfs}
\end{align}
where $A$ is the magnetic dipole-interaction and $Q$ the electric quadropole-interaction. The full one-particle Hamiltonian is then given by $H_\text{HFS}+ B \mu^z$, where the magnetic momentum operator is now given by $\boldsymbol{\mu}=\mu_B ( g_J \vec{J} +g_I \vec{I} )$.

The hyperfine Hamiltonian couples the state $\ket{m_J,m_I}$ with $\ket{m_J\pm 1,m_I \mp1}$ and $\ket{m_J\pm2,m_I\mp 2}$ and leaves the sum $m_J+m_I$ invariant. 
As we want to implement integer spin, we are interested in the limit of large magnetic fields where the eigenstates of the full Hamiltonian can be approximately described by quantum numbers $m_J$ and $m_I$. Due to the small nuclear magneton one can assume that the dipole-dipole-interactions in \cref{eq:dipoledipole} do not change $m_I$. 

In general, the hyperfine coupling affects not only the states with $m_J=\pm2$, but also all the other ground states. In particular, the states with $m_J=\pm1$ are also detuned, which results in the $D$-term in \cref{eq:heisenberg}. From the phase diagrams in \cref{fig:phasediagram} we know that the realization of the Haldane phase requires the detuning $D$ to be of the order of $J_{xy}$. 

Altogether, we are looking for eigenstates of $H_\text{HFS}+ B \mu^z$ that have a large overlap with $\ket{m_J=0,m_I}, \ket{m_J=\pm1,m_I}$ and $\ket{m_J=\pm2,m_I}$ for fixed $m_I$ with eigenenergies satisfying $|\Delta_+|,|\Delta_-|, |\Delta_+ +\Delta_-|\gg D \approx J_{xy}$. In the following we denote these eigenstates by $\ket{0'}$, $\ket{\pm1'}$ and $\ket{\pm2'}$. 
Numerically we find four different parameter sets with the discussed properties: 
\newline
\begin{center}
\begin{tabular} {l  c  r  }

$^{161}$Dy:  	&	$m_i=-5/2$ and		&	$B_0 = 1687.0$	G			\\
			&	$m_i=+3/2$ and		&	$B_0 = 3557.8$	G			\\						
$^{163}$Dy: 	&	$m_i=-5/2$ and		&	$B_0= 3476.5$ G 			\\
			&	$m_i=+1/2$ and		&	$B_0= 2094.2$ G		\\
			\\
\end{tabular}
\end{center}
Note that due to the admixing of other hyperfine states the states $\ket{\pm1'}$ couple to states different from $\ket{\pm2'}$. However, for all corresponding interactions the couplings are much smaller than the detunings, such that the transitions are strongly suppressed. 
In \cref{fig:perturbations} we exemplarily show the hyperfine spectrum of the ground state of $^{161}$Dy and the magnetic field dependence of the detuning and the static interaction $J_z$.

With the effective spin-1 systems occurring in fermionic dysprosium we are able to simulate the spin-1 XXZ-Heisenberg model. However, the resulting Hamiltonian is not exclusively described by \cref{eq:heisenberg} for two reasons. On the one hand additional terms occur due to the admixing of other hyperfine states, on the other hand virtual excitations into $\ket{\pm2'}$ cause Van-der-Waals like corrections. The resulting corrections are discussed in the following.

\subsubsection{Subleading corrections to the Hamiltonian}

\begin{figure}[tb]
\includegraphics[width= \linewidth]{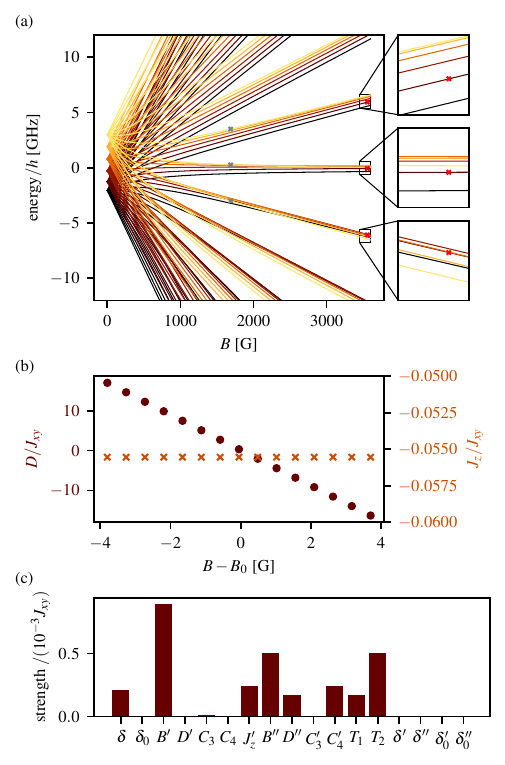}
\caption{Panel (a) shows the hyperfine splitting of the ground state of $^{161}$Dy. The crosses indicate the states that can be used to encode the spin-1 degrees of freedom. In the rest of this chapter we focus on the realization with $B_0=3557.8$ G. The insets show a zoom of the hyperfine structure in this region. In panel (b) we consider the magnetic field dependence of the detuning $D$ (dots) and the coupling $J_z$ (crosses) near $B_0=3557.8$. The lattice constant is set to $a=200$ nm and the angle between the magnetic field and the chain is $\theta=0$, which results in $J_{xy}/h \approx 90$ Hz. While the detuning can be adjusted via the magnetic field, the coupling $J_z$ is constant. In Panel (c) we show the coupling strengths of the perturbations introduced in \cref{eq:Hxy},  \cref{eq:Hz}, \cref{eq:H2} and \cref{eq:H3} for $^{161}$Dy and $B=3557.8$ G. Like the static interaction $J_z$, the perturbations are constant in the range of the magnetic fields considered in panel (b). }
\label{fig:perturbations}
\end{figure}

To write down the many-body Hamiltonian we determine the magnetic moments in the basis of the states $\ket{0'}$ and $\ket{\pm'}=\ket{\pm1'}$. The effective magnetic moments are of the form   
\begin{subequations}
\begin{align}
\mu^x&=\sqrt{2}^{-1} (d_1 \ket{+'}\bra{0'} + d_2 \ket{0'}\bra{-'} +\text{ h.c.})\\
\mu^y&=\sqrt{2}^{-1}(-i  d_1 \ket{+'}\bra{0'} +i  d_2 \ket{0'}\bra{-'} +\text{ h.c.})\\
\mu^z&= (m+\delta_z) \ket{+'}\bra{+'} -(m-\delta_z) \ket{-'}\bra{-'} +M \id
\end{align}
\end{subequations}
Inserting this into the dipole-dipole interaction in \cref{eq:dipoledipole} gives the XXZ-Heisenberg Hamiltonian in \cref{eq:heisenberg} with $J_{xy}=-\frac{\mu_0}{8 \pi a^3}(d_1^2+d_2^2)/2 (1-3 \cos^2(\theta))$ and $J_z=\frac{\mu_0}{4 \pi a^3} m^2 (1-3 \cos^2(\theta))$ and the additional correction terms discussed in the following. For a lattice spacing of $a=200$ nm and $\theta=0$ one obtains $J_{xy}/h \approx 90$ Hz. 

The difference of the couplings $d_1$ and $d_2$ leads to corrections of the form
\begin{align}
H_{xy}     =  \sum_{i < j} &\frac{1}{\left|i-j\right|^3}
                \left( \frac{\delta}{2} [ S^z_i S^+_i S^-_j S^z_j -  S^+_i S^z_i S^z_j S^-_j ]  \right. \notag \\ &\left.-\frac{ \delta_0}{2} [ S^z_i S^+_i S^z_j S^-_j + S^z_i S^-_i S^z_j S^+_j ]  + \text{h.c.} \right),
\label{eq:Hxy}
\end{align}
with $\delta=\frac{d_1^2-d_2^2}{d_1^2+d_2^2}J_{xy}$ and $\delta_0=-\frac{(d_1-d_2)^2}{d_1^2+d_2^2}J_{xy}$. For small differences $\epsilon=d_1-d_2 $ one finds $\delta\propto \epsilon$ and  $\delta_0\propto \epsilon^2$. Thus, $\delta_0$ is expected to be much smaller than $\delta$.

Furthermore, the fact that the state $\ket{0'}$ has a non-zero magnetic moment $M$ and the difference of the magnetic moments of the $\ket{\pm'}$ states cause terms of the form
\begin{align}
H_z= \sum_{i \neq  j}\frac{1}{\left|i-j\right|^3}& \left[    B^{\prime} \ S_i^z + D' \ (S_i^z)^2+C_3 \ S_i^z(S_j^z)^2 \right. \notag \\ &\left. +  C_4 \ (S_i^z)^2(S_j^z)^2   \right],
\label{eq:Hz}
\end{align}
where $B'=-\frac{4 Mm}{d_1^2+d_2^2}J_{xy}$, $D'=-\frac{4M\delta_z}{d_1^2+d_2^2}J_{xy}$, $C_3=-\frac{4m \delta_z}{d_1^2+d_2^2}J_{xy}$ and $C_4=-\frac{4\delta_z^2}{d_1^2+d_2^2}J_{xy}$.

In the previous section we showed that transitions to spin states with quantum numbers different from $0$ and $\pm1$ are strongly suppressed due to the large detunings of these states. However, in lowest order these virtual excitations give rise to two- and three-body interactions. For example starting with three atoms in the state $\ket{+',0',-'}$ there is a small probability that the system gets excited to the state $\ket{+2',-',-'}$. This state then evolves either back into the state $\ket{+',0',-'}$ or into the state $\ket{+',-',0'}$. To take these processes into account we use the Schrieffer-Wolff transformation. The resulting perturbations are
\begin{align}
H_2= \sum_{i \neq  j}\frac{1}{\left|i-j\right|^6} &\left[J_z^\prime\ S_i^zS_j^z      +B^{\prime \prime} \ S_i^z + D'' \ (S_i^z)^2 \right. \notag \\ &\left. +C_3' \ S_i^z(S_j^z)^2  + C_4'\ (S_i^z)^2(S_j^z)^2   \right],
 \label{eq:H2}
\end{align}
and 
\begin{align}
H_3=&\sum_{i\neq j \neq k}  \frac{1}{\left|i-k\right|^3} \frac{1}{\left|k-j\right|^3}  \left[ \left(T_1\ S^z_k +T_2\ (S^z_k)^2  \right)  S^+_iS^-_j \right. \notag \\
&+\left(\frac{\delta'}{2} S^z_k +\frac{\delta''}{2} (S^z_k)^2 \right) \left(  S^z_i S^+_i S^-_j S^z_j - S^+_i S^z_i S^z_j S^-_j \right) \notag \\ 
&+\left. \left(\frac{\delta_{0}'}{2}  S^z_k - \frac{\delta_{0}''}{2} (S^z_k)^2\right) \left( S^z_i S^+_i S^z_j S^-_j +S^+_i S^z_i S^-_j S^z_j\right)  \right]
 \label{eq:H3}
\end{align}
The two-body interaction contains terms similar to the $J_z$-part in \cref{eq:heisenberg} and the correction $H_z$ in \cref{eq:Hz}. The three-body Hamiltonian describes exchange interactions restricted by the fact, that a third atom has to be in the state $\ket{+'}$ or $\ket{-'}$. As in the derivation of $H_{xy}$ in \cref{eq:Hxy},  the terms containing $\delta',\delta'',\delta_{0}'$ and $\delta_{0}''$ originate from the different couplings $d_1$ and $d_2$. For $^{161}$Dy and $B=3557.8$ G the couplings $J_z'$,$B''$,$D''$,$C_3'$,$C_4'$, $T_1$, $T_2$, $\delta',\delta'',\delta_{0}'$ and $\delta_{0}''$ are exemplary shown in \cref{fig:perturbations}.

In principle virtual processes also allow transitions out of the spin-1 subspace. However, we numerically check, that all the states that couple via second order processes are out of resonance. 

Most of the corrections brake the symmetries that protect the Haldane phase. In particular the D$_2$ symmetry is broken by the terms proportional to $\delta$, $B'$,$C_3$,$B''$,$C_3'$, $T_1$,$\delta''$ and $\delta_{0}'$ and the time reversal symmetry is broken by $B'$,$C_3$,$B''$, $C_3'$. $T_1$,$\delta'$ and $\delta_0'$. 
From our results in \cref{fig:perturbations} we know that the dominating symmetry braking contributions in $H_{xy}$ and $H_z$ are $B'$ and $\delta$. As shown in the next section, the fractionalization of the edges can still be observed in the presence of these perturbations.

\subsubsection{Realization of the Haldane phase}

\begin{figure*}[tb]
\includegraphics[width= \linewidth]{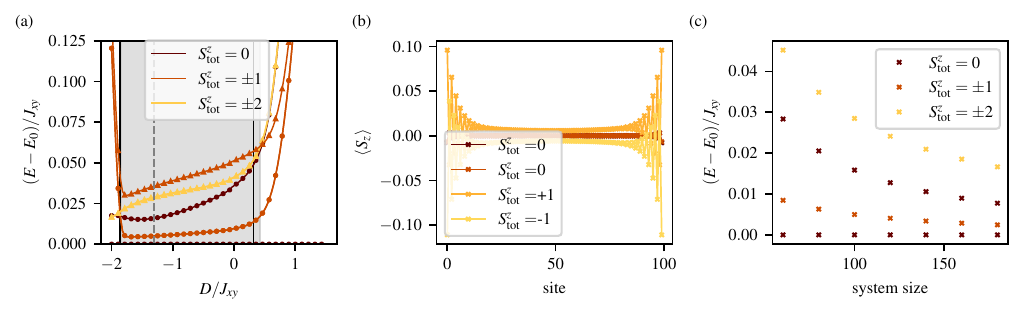}
\caption{Energy spectrum of the lowest energy state in the rotating wave approximation (a) and spin expectation values of the four ground states (b) at $D=-1.3 J_{xy}$ (dashed line in panel (a))  for a chain with 100 $^{161}$Dy atoms, lattice constant $a=200$ nm, $\theta=0$ and magnetic field $3557.47<B<3558.25$ (note that this implies $J_{xy}>0$  and $J_z=-0.0555 J_{xy}$). Panel (c) shows how the splitting of the ground state and the excitation gap change with the system size.  The energies and spin expectation values are calculated using DMRG with bond dimension $\chi=100$. The dashed vertical line in panel (a) indicates the point, that is analyzed further in panel (b) and (c). The straight vertical lines indicate the phase transition from the XY to the Haldane phase and from the Haldane phase to the large-D phase. The critical points are determined using the approach discussed in \cref{ap:boundaries}.  The region, where the four lowest states have magnetization $S^z_\text{tot}=\pm1$ and $S^z_\text{tot}=0$ ($\times 2$) are marked by the gray background. The corresponding states are marked with circles. The ground state degeneracy is lifted due to the finite size of the system. In panel (b) one clearly sees exponentially localized edge states. Due to the proximity of the phase boundary the edge states decay slow. In panel (c) we find that the splitting of the ground states decreases faster than the excitation gap. However, since we are very close to the boundary of the Haldane phase the splitting of the ground states is still of the order of the excitation gap for system size of $180$, that are much smaller than the characteristic length scale of the system near criticality.}
\label{fig:edgestates}
\end{figure*}

In this section we show that fermionic dysprosium atoms can be used to realize Haldane states that posses the characteristic spin-1/2 edge modes. Since the perturbations are the smallest for $^{161}$Dy and magnetic field $B \approx 3557.8 G$ we concentrate on this parameter regime. As shown in \cref{fig:perturbations}, the ratio of $J_z/J_{xy}$ is fixed and the only parameter that can be tuned arbitrarily is the detuning $D$. The regime that can be probed with fermionic Dysprosium is indicated by the gray dashed line in \cref{fig:phasediagram}. To find an optimal set of parameters we calculate the spectrum of the lowest energy states for a chain with 100 atoms along this line using repeated finite DMRG runs (see \cref{fig:edgestates}). In the Haldane phase we find that the four lowest states are the ones with magnetization $S^z_\text{tot}=\pm1$ and $S^z_\text{tot}=0$. However, the ground state degeneracy is lifted due to the finite size of the system and the overlap of the exponentially localized edge states. For larger system sizes the splitting of the ground states becomes smaller. For our example with 100 atoms the smallest splitting of the ground state energies occurs around $D \approx -1.3$, which implies, that the edge state decay length is the smallest in this region. 

Surprisingly, although the effective magnetic fields introduced in \cref{eq:Hz} and \cref{eq:H2} vary in space, the main effect of the perturbations on the low energy part of the spectrum corresponds to the splitting a global magnetic field would cause. This can be understood as follows. The effective magnetic field is constant in the bulk and differs only at the edges. In the Haldane phase, the lowest excited states are obtained by exciting one singlet in the bulk. Thus, the patterns of the spin expectation value of the lowest excited states are similar to the spin expectation values of the ground state and the effect of the effective magnetic field is the same for all low energy states. For our analysis of the energy spectrum in \cref{fig:edgestates} we absorb this global effective field into the external magnetic field $\mathbf{B}$.
 
In \cref{fig:edgestates} we show the spin expectation value of the four ground states at $D=-1.3$. As expected the spin expectation value alternates with the sites \cite{Miyashita1993}. However, the negative static interaction $J_z$ leads to an admixture of a ferromagnetic sub-pattern, which decays much slower. This large decay length is due to the fact that we are close to the edge of the Haldane phase. Furthermore, due to the  effective magnetic field the oscillations of the edge state with $S^z_\text{tot}=-1$ are stronger than the oscillations for $S^z_\text{tot}=+1$.

Note, that even in the presence of the symmetry breaking perturbations we observe a non-zero string order in the Haldane phase for system sizes of 100 atoms. However, this might be due to the finite size of the chain and does not necessarily imply that the string order survives the broken protecting symmetry.

\subsection{Bosonic Dysprosium with AC stark shifts}

\begin{figure}[tb]
\includegraphics[width= \linewidth]{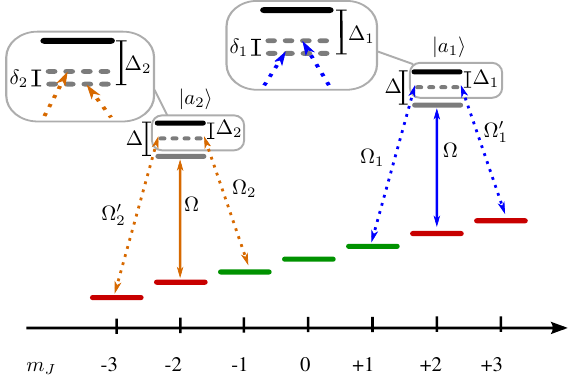}
  \caption{Level scheme for detuning the $\ket{\pm2}$ states (straight lines) using AC-stark shifts and tuning $J_z/J_{xy}$ (dashed lines). If the states $\ket{\pm3}$ are admixed to the effective spin-1 states an additional Stark shift has to be applied to shift the states $\ket{\pm4}$ out of resonance. To tune $J_z/J_{xy}$ one can also use $\ket{\pm2}$  instead of $\ket{\pm3}$. This has the advantage that fewer couplings are needed. However, a downside is that to obtain equally large $J_z/J_{xy}$ one has to admix more of the state $\ket{\pm2}$ such that the interaction strength is stronger decreased. }
  \label{fig:levelbosonic}
  \end{figure}

A second option to tune the $\ket{\pm 2}$ states out of resonance is to use of stark shifts. The general idea is the following. Each of the states $\ket{\pm 2}$ is coupled to an auxiliary state with coupling $\Omega$ and detuning $\Delta$ (see \cref{fig:levelbosonic}). For large detunings $\Delta \gg \Omega$  the energies of the states $\ket{\pm 2}$ can be approximated by $\epsilon_{\pm 2} = \pm 2 \mu_B g_J B -\Omega^2/\Delta$ and the energies one has to pay for leaving the Spin-1 subsystem are $\Delta_\pm=-\Omega^2/\Delta$.  Large detunings are not necessarily required, however if the states couple resonantly different couplings strengths $\Omega_\pm$ with $|\Omega_+ \pm \Omega_-| \gg V_{dd}$  are required. Note that virtual processes enable resonant transitions out of the spin-1 subspace with coupling strength of the order of $V_{dd}^2/\Delta_{\pm}$. However, for sufficiently large detunings $\Delta_\pm$ these processes are much slower than the time scales of an experiment.

So far we are limited to a fixed ratio of $J_z/J_{xy} \approx -0.0555$. However, mixing  $\ket{\pm 1}$ and for example $\ket{\pm 3}$ via Raman-type coupling allows to tune the ratio of $J_z/J_{xy}$, such that states deeply located in the Haldane phase can be realized.

We start with the admixing of the state $\ket{+ 3}$. The transitions between the ground states with quantum numbers $m_J=+1$ and $m_J=+3$ and an auxiliary state $\ket{a_1}$ are driven with couplings $\Omega_1$ and $\Omega_1'$. The lasers are detuned by $\Delta_1$ and $\Delta_1 -\delta_1$ (see \cref{fig:levelbosonic}). The system is then described by a Hamiltonian of the form 
\begin{align}
H=\begin{pmatrix}  0 & 0 &\Omega_1 \\ 0 &\delta_1 &\Omega_1' \\ \Omega_1 &\Omega_1' &\Delta_1 \end{pmatrix},
\end{align}
where we defined $\ket{+ 1}=(1,0,0)^T$, $\ket{+ 3}=(0,1,0)^T$ and $\ket{a_1}=(0,0,1)^T$. For $\delta_1 \ll \Delta_1\approx \Omega_1, \Omega_1' $ there exists a dark state $(\Omega_1', -\Omega_1,0)^T$ with energy $\epsilon \approx \delta_1 \Omega_1^2/(\Omega_1'^2+\Omega_1^2)$. The other two eigenstates 
are far detuned for $\Delta_1\approx \Omega_1, \Omega_1'$. 

Analog, $\ket{- 3}$ is admixed by coupling the ground states with quantum numbers $m_J=-1$ and $m_J=-3$ to an auxiliary state $\ket{a_2}$ with couplings $\Omega_2$ and $\Omega_2'$ and detunings $\Delta_2$ and $\Delta_2 -\delta_2$. The effective spin-1 states are then defined as 
\begin{subequations}
\begin{align}
\ket{+'} &=\frac{1}{\sqrt{\Omega_1^2+\Omega_1'^2}} (\Omega_1' \ket{+1}- \Omega_1 \ket{+3}) \\
\ket{-'} &=\frac{1}{\sqrt{\Omega_2^2+\Omega_2'^2}} (\Omega_2' \ket{-1}- \Omega_2 \ket{-3})\\
\ket{0'}&=\ket{0}. 
\end{align}
\end{subequations}
To ensure $\bra{+'} \mu^+ \ket{0'} = \bra{0'} \mu^+ \ket{-'}$ and $\bra{+'} \mu^z \ket{+'}=-\bra{-'} \mu^z \ket{-'} $ we require $\Omega_1/\Omega_1'=\Omega_2/\Omega_2':=\alpha$. The ratio of the couplings $\alpha$ allows us to vary 
\begin{align}
\frac{J_z}{J_{xy}}=-\frac{4}{72} \frac{(1+3 \alpha^2)^2}{1+\alpha^2}.
\end{align}
The detuning $D=(\delta_1+\delta_2) \alpha^2 /(1+\alpha^2)$ can be tuned via $\delta_1$ and $\delta_2$. The difference $\delta_1-\delta_2$ enters as a global correction of the magnetic field.

A suitable candidate for the stark shifts and the Raman scheme is the  684 nm transition from $4f^{10}6s^2$ $^5I_8$  to $4f^{9}5d 6s^2$ $^5I_8$ \cite{Schmitt2013}. Due to their different Lande factors the ground state and the excited state split differently in a magnetic field \cite{Martin1978}. As a consequence the transition frequency between a ground state and an excited state both with quantum number $m_J$ is $\omega_{m_J}=\omega_0+\Delta_{eg} m_J$ with $\Delta_{eg}/B=0.1539$ MHz/G. This allows to drive selected transitions between states with certain $m_J$- quantum numbers. As auxiliary states one could then use the excited states with quantum numbers $m_J=\pm2$. Note, that this realization requires significantly smaller magnetic fields than the realization with fermionic Dysprosium.

Due to the admixing of the excited state all ground states obtain a small decay rate  $\Gamma_g \propto \Omega^2/\Delta_{eg}^2 \Gamma$, here $\Omega$ represents the coupling strength of any of the six couplings introduced before. Furthermore, due to virtual processes different ground states are coupled with a coupling strength $ \propto \Omega^2/\Delta_{eg}$. However, this can be suppressed arbitrarily well choosing strong magnetic fields.

\section{Conclusion and Outlook}

In this work we present schemes for the realization of spin-1 XXZ-Heisenberg models using Dysprosium. The spin-1 degrees of freedom are encoded in the Zeemann sublevels of ground state Dysprosium. We provide two approaches to isolate spin-1 subsystems energetically. The first approach utilizes fermionic Dysprosium in a magnetic field, whereas in the second approach we make use of Stark shifts applied to bosonic Dysprosium.
We discuss the phase diagram of the resulting Hamiltonian and summarize the characteristic observables of these phases, that can be simulated using Dysprosium. In particular, we find that our setup allows for the realization of the Haldane phase and the observation of the characteristic edge states. Furthermore, we determine the subleading corrections that occur for the proposed scheme and show that the Haldane phase with spin fractionalization can be still observed in the presence of these perturbations.

The Haldane phase can be detected experimentally by measuring both string order parameters $\mathcal{S}^z_{ij}$ and $\mathcal{S}^x_{ij}$ and the  spin-spin correlations $\langle S_i^z S_j^z \rangle$, $\langle S_i^x S_j^x \rangle$, and $ \langle (S_i^+)^2 (S_j^-)^2 \rangle$. Then, the Haldane phase can be distinguished from other phases by observing that both string order parameters are non-zero combined with vanishing spin-spin correlations. In addition, measuring $\langle S_i \rangle$ locally allows to detect the edge states.

Due to its strong magnetic moment Dysprosium enables the simulation of spin-1 systems with exchange couplings as well as static couplings with dipolar character. The reachable parameter regimes differ from other proposals based on trapped ions \cite{Cohen2014,Cohen2015,Senko2015} and Rydberg atoms \cite{Moegerle2024}. Thus, our work allows for the exploration of new regions in the phase diagram of complex spin systems and is very well suited to study the influence of long-range interactions on quantum phases and phase transition.

\section{Acknowledgements}

We thank Jens Hertkorn for fruitful discussions. This work is supported by the German Rsearch Foundation (DFG) within FOR2247 under Grant No. Bu2247/1-2

\appendix

\section{Boundaries}
\label{ap:boundaries}

Here, we provide detailed information on the evalution of the critical lines shown in \cref{fig:phasediagram}. For all phase transitions except the anti-ferromagnet-Haldane transition we essentially follow reference \cite{Chen2003}.

For the latter we use the divergence of the entanglement entropy, which reflects a gap closing. The maximum of the entanglement entropy is determined for bond dimensions $8,16,24,32,50,64$ and near the anti-ferromagnet-Haldane-XY tricritical point for bond dimensions $24,32,50,64,100,200$. The critical values for $\chi \rightarrow \infty $ are obtained extrapolating with $c(\chi)=c_\infty+c_1\chi^{-c_2}$ \cite{Su2012}.

\begin{figure}[tb]
  \includegraphics[width= \linewidth]{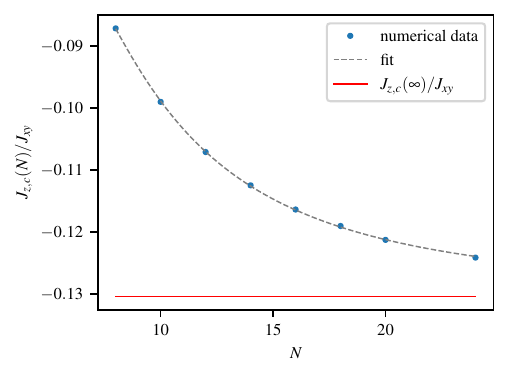}
  \caption{System size dependence of the critical value $J_{z,c}$ at $D=-1.2 J_{xy}$ for the XY-Haldane transition with $J_{xy}>0$. The red straight line indicates the value extrapolated for $N\rightarrow \infty$. The dashed line is a fit of the form $c(N)=c_\infty+c_1 N^{-2}+c_2 N^{-4}$.}
  \label{fig:boundary_over_site}
  \end{figure}

\begin{figure}[tb]
  \includegraphics[width= \linewidth]{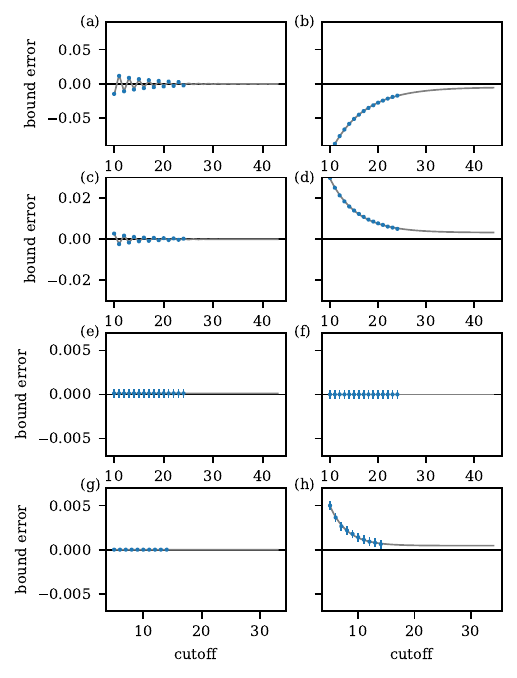}
  \caption{Deviation of the critical values for different cutoffs from the critical value without cutoff (zero-line) for different phase transitions. Panel (a) and (b) show the bound errors for the BKT-transition at  $D=0$ with system size $N=50$ for $J_{xy}>0$ (a) and $J_{xy}<0$, (b). Panel (c) and (d) show the bound errors for the XY1-XY2 transition for system size $N=50$, $J_z=-0.19$ and $J_{xy}>0$ (c) and $J_z=-0.1$ and $J_{xy}<0$ (d). Panel (e) and (f) show the bound errors for the Haldane-large-D transition for system size $N=50$, between $J_z=1$ and $D=0.65$ and $J_z=0.7$ and $D=0.88$ for $J_{xy}>0$ (e) and between $J_z=1.9$ and $D=1.3$ and $J_z=1.4$ and $D=1.8$ for $J_{xy}<0$ (f). Panel (g) and (h) show the bound errors for the FM-XY transition for system size $N=30$ at $D=0$ and $J_{xy}>0$ (g) and $D=-1.3$ and $J_{xy}<0$ (h). The gray lines show an extrapolation with $a (-1)^x\exp(-x/b)+c$ ($J_{xy}>0$) and $a\exp(-x/b)+c$ ($J_{xy}<0$). For large cutoffs the critical values obtained from extrapolation converge to the critical value without cutoff. The error bars represent the tolerance up to which we locate the critical values. For all critical values we used a bond dimension of 800. }
  \label{fig:cutoffdependence}
  \end{figure}

The transition from the XY phase to the Haldane phase and the large-D phase is a Berezinskii-Kosterlitz-Thouless (BKT) transition and the DMRG-method converges slowly in this region. Hence, finding the boundary comes along with several challenges. Here, we use the Level-spectroscopy method developed in \cite{Nomura1995,Kitazawa1996,Kitazawa1997}. Combining bosonization \cite{Timonen1985,Schulz1986,Giamarchi2007} and conformal field theory arguments \cite{Cardy1986} one can show that for periodic boundary conditions the energies of the lowest excited state with total spin $S^z_\text{tot}=4$ which is symmetric under spatial inversion ($\mathbf{S}_i \rightarrow \mathbf{S}_{N-i+1} $) and translation ($\mathbf{S}_i \rightarrow  \mathbf{S}_{i+1}$)  and the  lowest excited state with total spin $S^z_\text{tot}=0$ which is symmetric under spatial inversion and translation have to cross at the BKT-line. We determine the crossings of the excitations for $N=8,10,12,14,16,18,20,24$ ($J_{xy}>0$) and $N=20,24,28,32,36,40,44,50$ ($J_{xy}<0$) and bond dimension $\chi=400$ and extrapolate the critical values using $c(N)=c_\infty+c_1 N^{-2}+c_2 N^{-4}+c_3 N^{-6}$. For negative $J_{xy}$ the critical values near the XY-Haldane-large-D tricritical  point converge slowly and the exact position of the tricritical point is hard to determine. In \cref{fig:boundary_over_site} we show the system size dependence of the critical value $J_{z,c}$ at $D=-1.2 J_{xy}$ for the XY-Haldane transition. 

In contrary to the problems considered in \cite{Nomura1995,Kitazawa1996,Kitazawa1997,Chen2003}, our Hamiltonian includes long-range interactions. Thus, the question arises how to deal with periodic boundary conditions. The easiest is to introduce a cutoff that is smaller than half of the system size. However, as shown in figure \cref{fig:cutoffdependence} this converges slowly for some transitions. Furthermore, we find that one obtains the same critical values without taking into account a cutoff. Thus, we don't use a cutoff, which allows us to work with smaller system sizes.

To determine the critical line between the Haldane phase and the large-D phase we use twisted boundary conditions ($S^x_{N+1}=-S_1^x, S^y_{N+1}=-S_1^y,S^z_{N+1}=S_1^z$). In this case, the Haldane state is asymmetric under spatial inversion and time reversal while the large-D state is symmetric. Thus, when the system transitions from the large-D to the Haldane phase, the energy of the lowest asymmetric state becomes smaller than the energy of the lowest symmetric state \cite{Chen2003}. Again, we compare the critical values for different cutoffs in \cref{fig:cutoffdependence}. For $N=8,10,12,14,16,18,20,24$ and $\chi=400$ we determine the critical point using no cutoff and extrapolate the critical values using $c(N)=c_\infty+c_1 N^{-2}+c_2 N^{-4}$ \cite{Chen2003}. 

The boundary of the ferromagnetic phase is determined from the crossing of the lowest states with magnetization $S^z_{tot}=N$ and $S^z_{tot}=0$ \cite{Chen2003,Gong2016}. The energies of these states are calculated for system sizes $N=8,10,12,14,16,18,20,24$ and periodic boundary conditions without cutoff. The transition point is then obtained by extrapolating $c(N)=c_\infty+c_1 N^{-1}+c_2 N^{-2}$ \cite{Chen2003}.
The critical values rarely depend on the cutoff, and we obtain similar results with calculations where no cutoff is used \cref{fig:cutoffdependence}.   

For the boundary between the two different XY-phases we use the crossing of the two lowest excited states with magnetization $S^z_{tot}=1$ and $S^z_{tot}=2$ \cite{Chen2003}. As discussed in \cref{sec:phasediagram} for large negative $D$ the Hamiltonian in \cref{eq:heisenberg} maps to a spin 1/2-XXZ Heisenberg model with $\ket{\uparrow}=\ket{+1}$ and $\ket{\downarrow}=\ket{-1}$. Thus, the lowest excitation has magnetization $S^z_{tot}=\pm2$, which corresponds to flipping one atom from $\ket{+1}$ to $\ket{-1}$. However, for smaller $|D|$ the excitation with $S^z_{tot}=\pm1$ has  a smaller energy. We determine the crossing of these two excited states for system sizes $N=8,10,12,14,16,18,20,24$ and periodic boundary conditions without cutoff. The transition point is then obtained by extrapolating $c(N)=c_\infty+c_1 N^{-1}+c_2 N^{-2}$.

\bibliography{export}

\end{document}